\documentclass[longauth]{aa}
\usepackage{txfonts}
\usepackage{graphicx}
\usepackage{amssymb}
\usepackage{natbib}
\bibpunct{(}{)}{;}{a}{}{,}
\usepackage{color}
\def\FB {F_\mathrm{B}}
\def\FS {F_\mathrm{S}}
\def\DS {D_\mathrm{S}}
\def\DL {D_\mathrm{L}}

\def\tE {t_\mathrm{E}}
\def\ThE {\theta_\mathrm{E}}
\def\piE {\pi_\mathrm{E}}
\def\RS {R_\mathrm{*}}
\def\ThS {\theta_\mathrm{*}}
\def\rhoS {\rho_\mathrm{*}}
\def\msun {M_{\odot}}

\def\rsun {R_{\odot}}
\def\zsun {Z_{\odot}}

\def\teff {T_\mathrm{eff}}

\def\VIo {(V-I)_{\circ}}

\def\Io {I_{\circ}}
\def\Ri {R_\mathrm{I}}
\def\Ai {A_\mathrm{I}}
\def\logg {\log g}

\def\Mi {M_\mathrm{I}}
\def\chidof {\chi^2/\mathrm{d.o.f.}}
\newcommand\SEC[1]  {Sec.~\ref{#1}}
\newcommand\EQ[1]   {Eq.~(\ref{#1})}
\newcommand\Fig[1]  {Fig.~\ref{#1}}
\newcommand\Tab[1]  {Tab.~\ref{#1}}
\def\cf {\emph{cf.}~}
\newcommand \dix[1] {\times 10^{#1}}

\begin{document}
  \title{OGLE~2004--BLG--254: a K3~III Galactic Bulge Giant spatially resolved by a single microlens\thanks{Based on observations made at ESO, 073.D-0575A}}
  \titlerunning{OGLE~2004--BLG--254, a K3~III microlensed Galactic Bulge Giant}
  \authorrunning{ A.~Cassan, J.-P.~Beaulieu, P.~Fouqu\'e \emph{et al.}}
  \author{A.~Cassan\inst{1,2,3} \and J.-P.~Beaulieu\inst{1,3} \and P.~Fouqu\'e\inst{1,4} \and
    S.~Brillant\inst{1,5} \and M.~Dominik\inst{1,6} \and J.~Greenhill\inst{1,7} \and 
    D.~Heyrovsk\'y\inst{8} \and K.~Horne\inst{1,6} \and U.G.~J{\o}rgensen\inst{1,9} \and 
    D.~Kubas\inst{1,5} \and H.C.~Stempels\inst{6} \and C.~Vinter\inst{1,9} \and
    M.D.~Albrow\inst{1,12} \and D.~Bennett\inst{1,13} \and J.A.R.~Caldwell\inst{1,14,15} \and
    J.J.~Calitz\inst{1,16} \and K.~Cook\inst{1,17} \and C.~Coutures\inst{1,18} \and
    D.~Dominis\inst{1,19} \and J.~Donatowicz\inst{1,20} \and K.~Hill\inst{1,7} \and
    M.~Hoffman\inst{1,16} \and S.~Kane\inst{1,21} \and J.-B.~Marquette\inst{1,3} \and R.~Martin\inst{1,22} \and
    P.~Meintjes\inst{1,16} \and J.~Menzies\inst{1,23} \and V.R.~Miller\inst{12} \and K.R.~Pollard\inst{1,12} \and
    K.C.~Sahu\inst{1,14} \and J.~Wambsganss\inst{1,2} \and A.~Williams\inst{1,22} \and
    A. Udalski\inst{10,11} \and M.K. Szyma{\'n}ski\inst{10,11} \and M. Kubiak\inst{10,11} \and
    G. Pietrzy{\'n}ski\inst{10,11,24} \and I. Soszy{\'n}ski\inst{10,11,24} \and
    K. {\.Z}ebru{\'n}\inst{10,11} \and O. Szewczyk\inst{10,11} \and {\L}. Wyrzykowski\inst{10,11,25} }
  \institute{
    PLANET/Robonet Collaboration
    \and Astronomisches Rechen-Institut (ARI), Zentrum f\"{u}r Astronomie der Universit\"{a}t Heidelberg (ZAH), M\"{o}nchhofStr. 12­-14, 69120 Heidelberg, Germany
    \and Institut d'Astrophysique de Paris, UMR~7095 CNRS -- Universit\'{e} Pierre \& Marie Curie, 98bis Bd Arago, 75014 Paris, France 
    \and Observatoire Midi-Pyr\'{e}n\'{e}es, Laboratoire d'Astrophysique, UMR~5572, Universit\'{e} Paul Sabatier - Toulouse 3, 14 avenue Edouard Belin, 31400 Toulouse, France
    \and European Southern Observatory (ESO), Casilla 19001, Vitacura 19, Santiago, Chile
    \and University of St Andrews, School of Physics \& Astronomy, North Haugh, St Andrews, KY16~9SS, United Kingdom
    \and University of Tasmania, Physics Department, GPO 252C, Hobart, Tasmania 7001, Australia
    \and Institute of Theoretical Physics, Charles University, V Hole\v{s}ovi\v{c}k\'{a}ch 2, 180 00 Prague, Czech Republic
    \and Niels Bohr Institute, Astronomical Observatory, Juliane Maries Vej 30, DK-2100 Copenhagen, Denmark
    \and OGLE Collaboration
    \and Warsaw University Observatory. Al. Ujazdowskie 4, 00-478 Warszawa, Poland
    \and University of Canterbury, Department of Physics \& Astronomy, Private Bag 4800, Christchurch, New Zealand
    \and University of Notre Dame, Physics Department, 225 Nieuwland Science Hall, Notre Dame, IN 46530, USA
    \and Space Telescope Science Institute, 3700 San Martin Drive, Baltimore, MD 21218, USA
    \and University of Texas, McDonald Observatory, Fort Davis TX 79734, USA
    \and Dept Physics / Boyden Observatory, University of the Free State, Bloemfontein 9300, South Africa
    \and Institute of Geophysics and Planetary Physics, L-413, Lawrence Livermore National Laboratory, P.O. Box 808, Livermore, CA 94550, USA
    \and DSM/DAPNIA, CEA Saclay, 91191 Gif-sur-Yvette cedex, France
    \and Astrophysikalisches Institut Potsdam, An der Sternwarte 16, D-14482 Potsdam, Germany
    \and Technical University of Vienna, Dept. of Computing, Wiedner Hauptstrasse 10, Vienna, Austria
    \and Department of Astronomy, University of Florida, 211 Bryant Space Science Center, Gainesville, FL 32611-2055, USA
    \and Perth Observatory, Walnut Road, Bickley, Perth 6076, Australia
    \and South African Astronomical Observatory, P.O. Box 9 Observatory 7935, South Africa
    \and Universidad de Concepci\'{o}n, Departamento de F\'{\i}sica, Casilla 160-C, Concepci\'{o}n, Chile
    \and Jodrell Bank Observatory, The University of Manchester, Macclesfield, Cheshire SK11 9DL, United Kingdom
  }
  \date{ Received <date> / Accepted <date> }
  \abstract
      {}
      {	We present an analysis of \object{OGLE~2004--BLG--254}, a
	high-magnification ($A_{\circ} \simeq 60$) and relatively short
	duration ($\tE \simeq 13.2$ days) microlensing event in which the
	source star, a Bulge K-giant, has been spatially resolved by a
	point-like lens. We seek to determine the lens and source distance,
	and provide a measurement of the linear limb-darkening coefficients of
	the source star in the $I$ and $R$ bands. We discuss the derived values of the latter
	and compare them to the classical theoretical
	laws, and furthermore examine the cases of already published 
	microlensed GK-giants limb-darkening measurements. }
      { We have obtained dense photometric coverage of the event light curve
	with OGLE and PLANET telescopes, as well as a
	high signal-to-noise ratio spectrum taken while the source 
	was still magnified by $A \sim 20$, using the UVES/VLT spectrograph.
	We have performed a modelling of the light curve, including finite source 
	and parallax effects, and have combined spectroscopic and photometric 
	analysis to infer the source distance. A Galactic model for the mass and velocity 
	distribution of the stars has been used to estimate the lens distance. }
      { From the spectrum analysis and calibrated color-magnitude
	of the event target, we found that the source was a K3~III Bulge giant,
	situated at the far end of the Bulge. 
	From modelling the light curve, we have derived an angular size of
	the Einstein ring $\ThE \simeq 114\,\mu$as, and a relative
	lens-source proper motion $\mu = \ThE/\tE \simeq 3.1$ mas/yr.
	We could also measure the angular size of the source, $\ThS \simeq 4.5\,\mu$as,
	whereas given the short duration of the event,
	no significant constraint could be obtained from parallax
	effects. 
	A Galactic model based on the modelling of the light curve
	then provides us with an estimate of the lens distance, mass and velocity
	as $\DL\simeq 9.6$~kpc, $M\simeq 0.11\,\msun$ and 
	$v \simeq 145~\mbox{km}\,\mbox{s}^{-1}$ (at the lens distance) respectively.
	Our dense coverage of this event allows us to measure limb darkening of the 
	source star in the $I$ and $R$ bands.  We also compare previous measurements of linear
	limb-darkening coefficients involving GK-giant stars with
	predictions from ATLAS atmosphere models. We discuss the case of
	K-giants and find a disagreement between limb-darkening
	measurements and model predictions, which may be caused by
	the inadequacy of the linear limb-darkening law. }
      {}
\maketitle
      
\section{Introduction} \label{sec:intro}

The microlensing technique is one of a few methods (together with
interferometry, eclipsing binaries and transiting extra-solar planets)
that can be used to measure brightness profiles and limb-darkening coefficients
of stars at distances exceeding a few kpc. The apparent stellar disk is a projection of the
near stellar hemisphere and therefore shows an axisymmetric variation
of the observed brightness as the projection maps the variation of the
emergent radiation with depth. The continuum is on average formed at
larger depth at the disk center and at smaller depth at the limb
\citep[Eddington-Barbier effect, see
e.g.:][]{WM1994,Dim95,Sasselov1997, HSL2000,Heyrovsky2003}. The
microlensing method requires the source to transit a region with a
large magnification gradient over its face, as present in the vicinity
of caustics.  While a single lens creates a point-like caustic at its
angular position, binary lenses create extended caustic patterns
formed of lines (fold caustics) which merge at cusps.

PLANET observations of the microlensing event \object{MACHO~1997--BLG--28}
\citep{Albrow1999b} containing a cusp passage, constituted the first
limb-darkening measurement of a Bulge giant. The majority of
subsequent limb-darkening measurements such as
MACHO~1998--SMC--1 \citep{Albrow1999a, Afonso2000}, \object{MACHO~1997--BLG--41}
\citep{Albrow2000b}, \object{OGLE~1999--BLG--23} \citep{Albrow2001a},
\object{EROS--2000--BLG--5} \citep{Fields2003,An2002}, \object{OGLE~2002--BLG--069}
\citep{Cassan2004,Kubas2005,Thurl2006} resulted from fold-caustic passages. 
The limb-darkening measurement in the solar-like star
\object{MOA~2002--BLG--33} \citep{Abe2003} constituted a very special case with
the source enclosing several cusps of the caustic at the same time.
In the single lens case,  the extended size of the source star 
has a significant effect on the light curve if 
the angular source size is
of the order or larger than the angular separation between 
source center and lens point-like
caustic. However, only a small fraction of microlensing events provide
angular separations that are small enough, and so far very few cases
have been observed \citep{Alcock1997,Yoo2004,Jiang2004}.  For events
showing evidence of this effect, the limb darkening of the source star
can be determined, and constraints on the physical properties of the
lens can be derived using estimates of its distance, spectral type and
event model parameters \citep{Gould1994, Witt1995, Dim95, HSL2000,
Heyrovsky2003}.
 
This work treats the case of OGLE~2004--BLG--254, a high magnification
event from a point-like lens showing extended source effects, for
which a dense photometric follow-up was performed at four PLANET
observing sites, making it one of the best observed events of this
kind to date.  Using high-resolution spectra collected with UVES
(VLT), we derive the characteristics of the source star, a K3
giant. We search for a suitable single-lens limb-darkened source
microlensing model, which we use together with the properties derived
from the high resolution spectra to discuss the source center-to-limb
variations and constraints on the lens properties.  We compare our
limb-darkening measurements to previous observations of such stars,
especially to EROS--2000--BLG--5 for which a large discrepancy was
found between observations and atmosphere models.  We also note that
the derived characteristics provide the basic information necessary
for a forthcoming study of element abundance in the Bulge giant source
star of OGLE~2004--BLG--254.

\section{Photometric and spectroscopic observations}

\subsection{Photometric monitoring}  \label{sec:data}

The OGLE-III Early Warning System (EWS) \citep{Udal03} discovered and
alerted the Bulge event OGLE~2004--BLG--254 ($\alpha=$~17h56m36.20s,
$\delta=-32\degr 33\arcmin 01\farcs 8$ (J2000.0) and $l=358.09\degr$, 
$b=-3.87\degr$) on May 17, 2004, from observations carried out with 
the 1.3 m Warsaw Telescope at the Las Campanas Observatory (Chile).

The PLANET collaboration started its photometric observations on June
8 which form the basis for our analysis and consist of data from
4~different telescopes being part of the PLANET network: the Danish
1.54m at ESO La Silla (Chile), the Canopus 1m near Hobart (Tasmania),
the Elizabeth 1m at the South African Astronomical Observatory (SAAO)
at Sutherland and the Rockefeller 1.5m of the Boyden Observatory at
Bloemfontein (both in South Africa).  Every 30 minutes the data from
the different sites are uploaded to a central computer in Paris, where
data are combined and fitted automatically, and light curves are made
publicly available.

The event was also monitored by the $\mu$FUN collaboration from Chile
(1.3m telescope at the Cerro Tololo Inter-American Observatory) and
Israel (Wise 1m telescope at Mitzpe Ramon). Although we did not make
use of them for our modelling, we checked they were consistent with our data sets.

The data we collected on OGLE~2004--BLG--254
showed a rise in apparent brightness by $2.85$ mag above baseline 
on June 9, 8:10 UT. These pre-peak
data and adequate modelling predicted a peak to occur on June 10,
6:35$_{-30 \mathrm{min}}^{+20 \mathrm{min}}$ UT, at a rather
uncertain, but in any case large, magnification of $A_{\circ} \simeq
80_{-30}^{+70}$.  Events of this type harbour an exceptional potential
for the discovery or exclusion of extra-solar planets as well as for
the study of stellar atmospheres and might provide an opportunity for
measuring the mass of the lens star.

On June 10 at 12:45 UT, a public alert was issued by PLANET reporting
that data collected on OGLE~2004--BLG--254 at the SAAO 1.0m between
June 9, 18:50 UT and June 10, 4:40 UT, at the Danish 1.54m on June 10
between 2:20 UT and 10:05 UT, as well as OGLE data obtained on June 10
between 3:50 UT and 9:55 UT, revealed the extended size of the source
star, and the crossing time of the disk was estimated to be around
$16$ hours.  The peak was passed around June 10, 7:40 UT at $4.35$ mag
above baseline, at a magnification of $A \sim 55$.

\subsection{Blending of the source star}  \label{sec:254blend}

There are two nearby stars on the eastern side of
OGLE~2004--BLG--254. Their calibrated OGLE magnitudes and colours are:
$I=16.31$, $(V-I)=2.30$ for the star located $0.23\arcsec$ north and 
$1.41\arcsec$ east, and $I=16.69$, $(V-I)=1.55$ for the star located 
$0.52\arcsec$ north and $3.00\arcsec$ east.

Thanks to its relative brightness, the source star of
OGLE~2004--BLG--254 can be found in recent infrared surveys, such as
DENIS \citep{Epchtein1994} or 2MASS \citep{Skrutskie1997}. However,
due to the large pixel size of these surveys ($1$ to $3 \arcsec$) and
the proximity of a companion of similar colour, the infrared measurements 
probably correspond to a blend of these stars: DENIS measured a ``star'' 
at $1.0 \arcsec$ E and $1.5 \arcsec$ S from OGLE~2004--BLG--254, 
with $I = 15.81 \pm 0.09$, $J = 14.02 \pm 0.12$, 
$K_{\mathrm s} = 13.12 \pm 0.16$, while 2MASS measured the same ``star'' 
at $1.3 \arcsec$ E and $1.0 \arcsec$ S, with $J = 14.135 \pm 0.046$, 
$H = 13.326 \pm 0.048$ and $K_{\mathrm s} = 12.969 \pm 0.045$. 2MASS 
quality flags are optimal, and blend and
confusion flags are not activated. The DENIS $I$ magnitude is
therefore $0.78$ mag brighter than the OGLE calibrated measurement (\cf \SEC{sec:CMDcalib}),
which supports our hypothesis that DENIS measured a blend of the two
nearby stars of similar magnitudes.

\subsection{UVES spectroscopy} \label{sec:UVES}

On June 11, 2004, between 00:24 and 00:52 UT (HJD $= 2,453,167.54159$ at mid-exposure),
we obtained a high-resolution spectrum  of
OGLE~2004--BLG--254 using the 
UVES spectrograph mounted at the Nasmyth focus of Kueyen (VLT second unit).
At this time, the source flux was magnified 
by the gravitational lens by a factor $20$, making the VLT
equivalent to a $\sim 37$ m diameter telescope.
The spectrum was taken at one of the standard red settings centered on $5800$ \AA,
which covers the spectral domain $4780 - 6835$~\AA\
at a mean resolution of $\sim 40,000$, and typical S/N of $100$.

The data were reduced using version 2.1 of the UVES context of the MIDAS data 
reduction software.
The raw science data were first bias-substracted and then wavelength-calibrated
order by order using a Thorium-Argon lamp.
The position of the science spectrum was determined in each order and an 
optimum extraction was used to obtain 1D science spectra where the sky lines 
had been removed.
An Halogen lamp was used to obtain flat-field spectra, which were 
bias-substracted and combined, then wavelength-calibrated. The spectra
were then flat fielded order by order, and merged to obtain the final data.
The detail analysis of the final spectrum is detailed in \SEC{sec:fundamentalp}.

\section{Modelling of the light curve} \label{sec:models}

\begin{figure}[!h]
  \begin{center}
    \includegraphics[width=9cm]{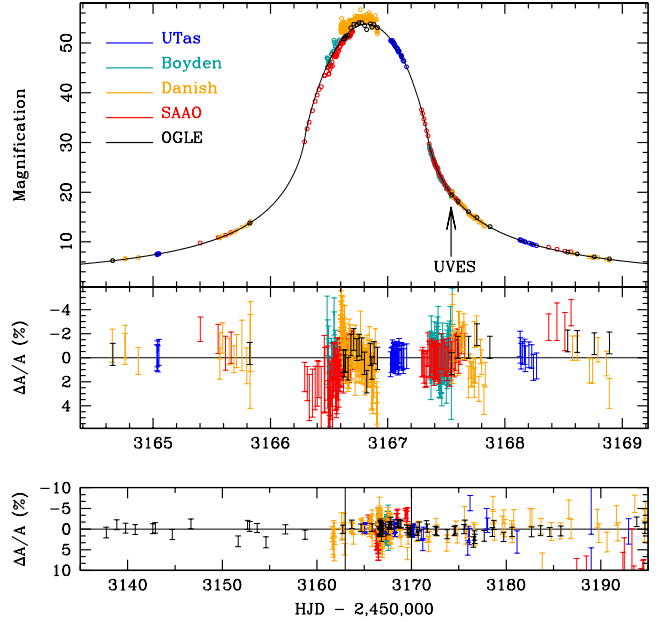}
    \caption{The upper panel shows the photometry of the microlensing
      event OGLE~2004--BLG--254 near its peak, on June 10, 2004, observed by
      four PLANET sites, Danish~1.54m, UTas~1m, Perth~0.6m, Boyden~1.5m and
      SAAO~1.0m and OGLE. Since the shape of the magnification curve depends
      on the value of each linear limb-darkening coeffecient, different for 
      each telescope, we choose the common best value of UTas and OGLE
      in \Tab{tab:fitparameters} to plot (solid line) the best-fitting point-lens,
      limb-darkened extended source model around the peak region.
      The residuals for each site are displayed by assuming their
      own best linear limb-darkening coefficients.
      The lower panel shows the residuals of the complete set of data (the
      two vertical lines indicate the peak region displayed above). The UVES 
      spectrum was taken at MHJD$= 3167.54159$ at mid-exposure (vertical arrow).   }
    \label{fig:lc}
  \end{center}
\end{figure}

The photometric data collected on OGLE~2004--BLG--254 (see \Fig{fig:lc})
clearly show that
the light curve is affected by extended source effects.

\subsection{Extended-source formalism}\label{sub:espl}

With $D_\mathrm{S}$ denoting the source distance and $x=\DL/\DS$ 
the fractional distance of the lens, of mass $M$, the angular Einstein radius reads
\begin{equation}
  \ThE = \sqrt{\frac{4GM}{c^2 \DS}\ \frac{1-x}{x}}.
\end{equation}
which is a characteristic scale of microlensing.
For a point source situated at a projected angular separation $u \ \ThE$ from the lens, 
the magnification function is given by
\begin{equation}
  A_\mathrm{PSPL}(u) = \frac{u^2 +2}{u\ \sqrt{u^2+4}}.
\end{equation}
The magnification $A_\mathrm{ES}$ of an extended limb-darkened source of
angular radius $\rhoS \ \ThE$ is obtained by integrating $A_\mathrm{PSPL}$
over the source disk.
With $I(r)$ being the
brightness profile of the source, normalized to unit flux (Sect.~\ref{sub:modelling}),
\begin{equation}
  \begin{array}{rl}
    A_\mathrm{ES}(u|\rhoS,I) &= \displaystyle \int\limits_0^{2\pi}\int\limits_0^1 
    I(r)\ A_\mathrm{PSPL}(\eta)\ r\ \mathrm{d}\varphi\ \mathrm{d}r,  \\
    \mbox{where}\qquad \eta &= \displaystyle \sqrt{(\rhoS\,r)^2 - 2\,u\,r\,\rhoS \cos \varphi + u^2 }.
  \label{eq:magext}
  \end{array}
\end{equation}

For a uniformly bright source, \citet{WM1994}
have derived a semi-analytic expression of $A_\mathrm{ES}$, involving elliptic
integrals. However, based on the fact that in a usual microlensing event toward 
the bulge of the Milky Way, extended source effects are only prominent for 
small lens-source angular separations ($u \ll 1$),
where $A_\mathrm{PSPL}(u) \simeq u^{-1}$, \citet{Gould1994} found that the 
extended-source magnification factorizes as:
\begin{equation}
  A_\mathrm{ES}(u|\rhoS) \simeq A_\mathrm{PSPL}(u) \times B_0\left(\frac{u}{\rhoS}\right).
\label{eq:yoo}
\end{equation}
The second factor can be expressed by the semi-analytical formula 
$B_0(z)=\frac{4}{\pi}\,z\,E\left[\arcsin \min\left(1, \frac{1}{z}\right),z\right]$ \citep{Yoo2004} ,
where $E$ is the incomplete elliptic integral of the second kind.
By separating the $u$ and $z=u/\rhoS$ parameters, this formula allows easy
discretization and fast computation of extended-source effects.

For non-uniform profiles (e.g. power-law limb-darkening
models or tabulated profiles from stellar atmospheric models), no such simple expressions are known. 
Different strategies have been proposed: 
\citet{WM1994} use the uniform source magnification and its derivative in
a one-dimensional integral; 
\citet{Heyrovsky2003} first calculates analytically the angular integral of the magnification,
so that a single integral involving the (radially dependent) 
brightness of the source remains. Finally, \citet{Yoo2004} give expressions of
$B_1$-and $B_{1/2}$- functions, related to the linear and square root 
limb-darkening laws, respectively (with the same approximations as for 
\EQ{eq:yoo}, to be numerically integrated.

When applying this formalism to OGLE~2004--BLG--254, we find
that the maximal relative discrepancy
between the exact magnification and its approximation related to
\EQ{eq:yoo} is less than $0.05\ \%$ for a linear limb-darkening law,
which is well within the typical data error bars.
We therefore use \EQ{eq:yoo} to derive our limb-darkening coefficients.

\subsection{Event model parameters} \label{sub:modelling}

In the photometric analysis of the event, both PLANET and OGLE data are used.
For each PLANET observation site, we have applied a cut on seeing
which only removes very unreliable points; we restricted the complete OGLE data to the ones
collected after HJD'~$=3050.0$ (which is large enough to derive the baseline magnitude).
OGLE provides $128$ data points, SAAO $114$, UTas $64$, Boyden $75$ and Danish $231$, 
for a total of $5$ observing sites and $612$ measurements.
Data reduced with our DoPhot-based pipeline
underestimate photometric errors (e.g. for bright magnitudes, error estimates
may reach $10^{-3}$, which is unrealistic). Comparison with the scatter of 
the non-variable stars suggests that the errors are underestimated by  about $20 \%$. 
Based on PLANET experience of PSF photometry used here, we rescale the errors as 
$\sigma_\mathrm{resc}^2 = \left(1.2\ \sigma_\mathrm{data}\right)^2 + (0.01)^2$. 
As the source star is a K3 giant, we first check 
if the fluctuation in the OGLE baseline magnitude can be due to intrinsic periodic activity
(which could easily be included into the models).
A power spectrum of 115 baseline data points coming from OGLE does not show
evidence for modulation greater than $\sigma_\mathrm{mod}=0.015$. 
We therefore attribute the residuals in the baseline
to non-periodic variability or observational noise.

As expected, a single-lens point-source model results in a high $\chidof$ ($\sim 31$), 
which obviously excludes this possibility.
On the other hand, a uniform extended source model provides us with a first working set of parameters
that fit the data well. However, at this stage, the residuals of the fit show some symmetric trends around the peak
of the light curve that indicate limb-darkening effects.
We thus add linear limb darkening to the source model. This is described here by
\begin{equation} 
  I(r) = \frac{1}{\pi} \left[ 1 - \Gamma \left(1-\frac{3}{2}\ \sqrt{1-r^2}\right) \right],
\end{equation}
where $I(r)$ is normalized to unit total flux.
The relation between $\Gamma$ and the more 
commonly used parameter $a$, defined by 
\begin{equation}
  I(r)\propto 1 - a \left( 1-\sqrt{1-r^2}\right),
\end{equation}
is given by
$a = \frac{3\,\Gamma}{2+\Gamma}$.

The parameterization involves basic microlensing parameters: 
$t_{\circ}$ (time of closest approach), $u_{\circ}$ (impact parameter), $\tE$ (crossing time of the Einstein ring radius), 
the source size $\rhoS$ (in units of $\ThE$) and two annual parallax parameters $\piE$
and $\psi$ (see Sect. \ref{sec:const}) common to all data sets, and specific parameters for each site:
baseline and blending magnitudes, as well as the linear limb-darkening coefficients $\Gamma$.
The values found for the latter are more fully discussed in \SEC{sub:LDmesure}.
Errors on the best-fit parameters were obtained by 
Monte-Carlo simulations:
we generate 500 sets of noisy light curves for each of the five observing sites using the obtained 
best-fit parameters, based on the rescaled error bars of the data; then,
from the distribution of the obtained values we determine the (non-Gaussian) $68.3\ \%$ 
confidence intervals for each parameter. The best corresponding set of parameters fitting 
the data and their errors is given in \Tab{tab:fitparameters}.

\begin{table}[!h]
 \begin{center}
    \caption{Parameters for the best-fitting point-lens limb-darkened extended source 
      models, using linear limb-darkening law (one LLDC coefficient per data set).
      The set of data 
      contains $612$ measurements coming from $4$ PLANET sites (Danish~1.54m, 
      UTas~1m, Boyden~1.5m and SAAO~1.0m) and from OGLE. We also report the 
      $\chi^2$ value.}
    \begin{tabular}{p{3cm}l}
      \hline
      ~~~ Parameters  & Value \& Error ~~~\\
      \hline\hline
      & \\
      ~~~ $t_{\circ}$ (days)\dotfill & $3166.8194 \pm 0.0002$ ~~~\\
      ~~~ $u_{\circ}$\dotfill & $4.60^{+0.76}_{-0.86}~\times10^{-3}$ ~~~\\
      ~~~ $\tE$ (days)\dotfill & $13.23^{+0.04}_{-0.05}$ ~~~\\
      ~~~ $\rhoS$\dotfill & $4.00^{+0.00}_{-0.02}~\times10^{-2}$ ~~~\\
      ~~~ $\left.\Gamma_I\right|_\mathrm{SAAO}$ / $\left.a_I\right|_\mathrm{SAAO}$ \dotfill & $0.35 ^{+0.02}_{-0.05}$ / $0.45 ^{+0.03}_{-0.06}$ ~~~\\
      ~~~ $\left.\Gamma_I\right|_\mathrm{UTas}$ / $\left.a_I\right|_\mathrm{UTas}$ \dotfill & $0.43 ^{+0.03}_{-0.06}$ / $0.53 ^{+0.03}_{-0.06}$ ~~~\\
      ~~~ $\left.\Gamma_I\right|_\mathrm{Boyden}$ / $\left.a_I\right|_\mathrm{Boyden}$ \dotfill & $0.58 ^{+0.00}_{-0.03}$ / $0.68 ^{+0.00}_{-0.03}$ ~~~\\
      ~~~ $\left.\Gamma_I\right|_\mathrm{OGLE}$ / $\left.a_I\right|_\mathrm{OGLE}$ \dotfill & $0.43 ^{+0.04}_{-0.06}$ / $0.53 ^{+0.04}_{-0.06}$ ~~~\\
      ~~~ $\left.\Gamma_R\right|_\mathrm{Danish}$ / $\left.a_R\right|_\mathrm{Danish}$ \dotfill & $0.61 ^{+0.03}_{-0.06}$ / $0.70 ^{+0.03}_{-0.05}$ ~~~\\
      ~~~ $\left.\FB/\FS\right|_\mathrm{SAAO}$\dotfill & $0.25 \pm 0.02$  ~~~\\
      ~~~ $\left.\FB/\FS\right|_\mathrm{Danish}$\dotfill & $0.05 \pm 0.01$ ~~~\\
      ~~~ $\left.\FB/\FS\right|_\mathrm{UTas}$\dotfill & $0.55 \pm 0.01$  ~~~\\
      ~~~ $\left.\FB/\FS\right|_\mathrm{Boyden}$\dotfill & $1.15 \pm 0.2$ ~~~\\
      ~~~ $\left.\FB/\FS\right|_\mathrm{OGLE}$\dotfill & $0.00 \pm 0.0$ ~~~\\
      ~~~ $\chidof$\dotfill & $1326/593$ ~~~\\
      & \\
      \hline
    \end{tabular}
    \label{tab:fitparameters}
 \end{center}
\end{table}

The blending fraction significantly varies from one PLANET site to another. 
This is a consequence of the fact that the source has two very close 
neighbours, one of them at a distance of $1.4''$ as stated in \SEC{sec:254blend}.
Our measured blendings are consistent with the different telescopes CCD resolutions. Moreover,
the low value obtained with the Danish 1.54m ($\simeq6\%$) favors a low blend flux
from the lens star.

Based on this model, we find the transit time to be $\simeq 1.05$~day and the date (where we write $t={\rm HJD}-2,450,000$) 
at which the lens starts to transit 
the source disk, $t_{\rm entry} \simeq 3166.2937$, as well as the exit time $t_{\rm exit} \simeq 3167.3451$.
One can note the UVES spectra (\SEC{sec:UVES}) were taken after the end of the transit, and so only weak differential magnification over the source disk 
is expected \citep{HSL2000}.

\section{Nature of the source star} \label{sub:srceparam}

Measuring the relative proper motion and velocity at the lens distance requires 
a determination of the 
distance and radius of the source star. This can be achieved 
by determining the apparent magnitude at baseline, the effective temperature, 
the gravity, and the amount of interstellar absorption toward 
the source. We seek to measure these quantities by use of a combination of 
the obtained photometry and spectroscopy.

\subsection{Source star calibrated colour-magnitude diagram} \label{sec:CMDcalib}

\begin{figure}
  \begin{center}
    \includegraphics[width=9cm]{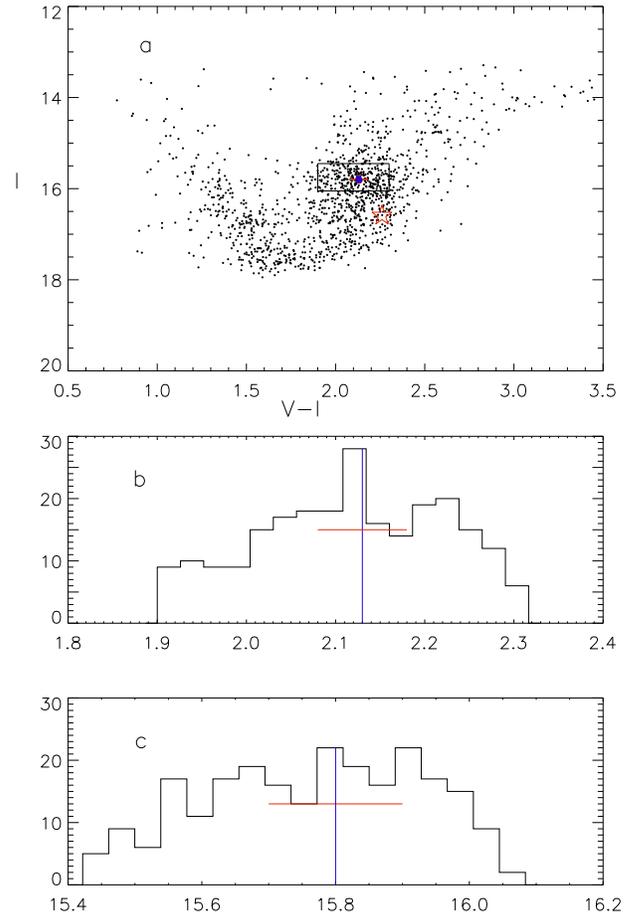}
    \caption{(a) Calibrated $(V-I), I$ colour-magnitude diagram of the 
      field around OGLE~2004--BLG--254 as obtained with the OGLE telescope at
      Las Campanas. The target is plotted as a red open star. The mean position of the
      red giant clump is the blue dot, as determined by the color histogram (b) and
      magnitude histogram (c) of the stars inside the black box of panel (a). Mean positions
      and uncertainties are the blue and red lines respectively.}
      \label{fig:colmag}
      \label{fig:colmag2}
  \end{center}
\end{figure}

\Fig{fig:colmag} shows the calibrated OGLE $(V-I)$ versus $I$
colour-magnitude diagram of the field of the source/lens. The calibrated
magnitude and colour of the target are $I=16.59 \pm 0.05$ and 
$(V-I)=2.26 \pm 0.08$. A box encompasses stars 
belonging to the red giant clump (RC), and the two histograms give a calibrated mean 
position of the clump of $I^{\rm RC} = 15.80 \pm 0.10$ and $(V-I)^{\rm RC} = 2.13 \pm 0.05$.
The shift in position of our target with respect to the mean red clump position is then
$\Delta I=0.79 \pm 0.11$ and $\Delta(V-I)=0.13 \pm 0.09$.

For the absolute magnitude of the clump, we adopt the Hipparcos position
as given by \citet{Stanek1998}, $\Mi=-0.23 \pm 0.03$. The mean
Hipparcos clump colour of ${\VIo}^{\rm RC}=1.00 \pm 0.05$ is adopted,
assuming any difference with the Baade's Window mean colour to be due
to the anomalous extinction law in the Galactic Bulge, as first explained by 
\citet{Popowski2000}.

The Galactic Center distance modulus from \citet{Eisenhauer2005} is adopted
as $\mu=14.41 \pm 0.09$, but a slightly larger distance is adopted for the
red clump in this field, due to its negative longitude and the bar geometry
\citep[see, e.g.:][]{Stanek1997}.
The difference in distance is based on mean clump position in different
regions from \citet{Sumi2004}, and amounts to $\Delta \mu=0.21 \pm 0.10$, 
giving an adopted distance to the red clump of this field of 
$\mu=14.62 \pm 0.13$. Therefore, the mean dereddened magnitude of the clump
is ${\Io}^{\rm RC} = 14.39 \pm 0.14$. Comparison with its calibrated apparent 
magnitude and colour gives an estimate of reddening parameters as 
$\Ai=1.41 \pm 0.17$ and $E(V-I)=1.13 \pm 0.07$, the latter in very good 
agreement with Sumi's measurement of $E(V-I)=1.17$ for the nearby OGLE-II 
field BUL-SC25. From these two independently determined
values follows the total-to-selective absorption ratio 
$\Ri=\Ai/E(V-I)=1.25 \pm 0.14$, intermediate between standard reddening law value 
(1.5) and Sumi's anomalous mean value for the Galactic Bulge (0.964).

We furthermore use \citet{Girardi2002} theoretical isochrones to determine
the mean metallicity of the red clump, and by adopting an age of $10$~Gyr (typical age of red clump
giants in the Bulge) we find $Z^{\rm RC}\simeq 0.008$.

Finally, from the adopted position of the clump and the shift of the target, 
we obtain a dereddened magnitude and colour of the target as 
$\Io=15.18 \pm 0.18$ and $\VIo=1.13 \pm 0.11$.

The comparison of $\VIo$ with the values from the 
grid of model atmosphere colours by \citet{Buser1992} provides an estimate
of the photometric temperature of the source star.
First, we note that typical changes in the theoretical value of $\VIo$
for a change in metallicity by a factor of two, or a change
in $\logg$ by an order of magnitude, are about $5\dix{-3}$ (for fundamental
parameters in the range of values of interest in the present study).
In comparison, a change in the value of $\teff$ by $100$~K gives rise 
to a 10 times larger change in $\VIo$. The value of $\VIo$ is therefore
primarily a measure of the effective temperature of the star.
Interpolating in the values of $\VIo$ from the grid of 
\citet{Buser1992},
gives us the best fit effective temperature of the source star 
from the photometry alone, as $\teff^\mathrm {phot} = 4500 \pm 250\,\mbox{K}$.
The quoted uncertainty reflects solely from the 
uncertainty in the estimate of $\VIo$.

\subsection{Source star fundamental parameters from spectroscopy} \label{sec:fundamentalp}

The position of the source star (and lens) is in the direction of the 
Sagittarius dwarf galaxy, and it could either be 
a member of the dwarf galaxy or of the Galactic Bulge. 
The line positions of the observed spectrum have a general offset 
of $+134$ km/s compared to laboratory data, which is also consistent 
with the star being in either the Sagittarius dwarf galaxy or the Galactic Bulge.
The mean chemical abundance [\element{Fe}/\element{H}] ranges between $-0.8$ and $-1.2$ 
in Sagittarius, 
and around $-0.1$  in the Galactic Bulge \citep{Fulbright05}. 
For the analysis of the UVES spectrum, we have therefore
computed a grid of model atmospheres with $\teff$ between $4000$~K and 
$4600$~K, $\logg$ between $0.0$ and $3.0$, and scaled solar abundances
with [\element{Fe}/\element{H}] between $-2.0$ and $+0.5$, and corresponding synthetic spectra.

\begin{figure*}
  \includegraphics[width=16cm]{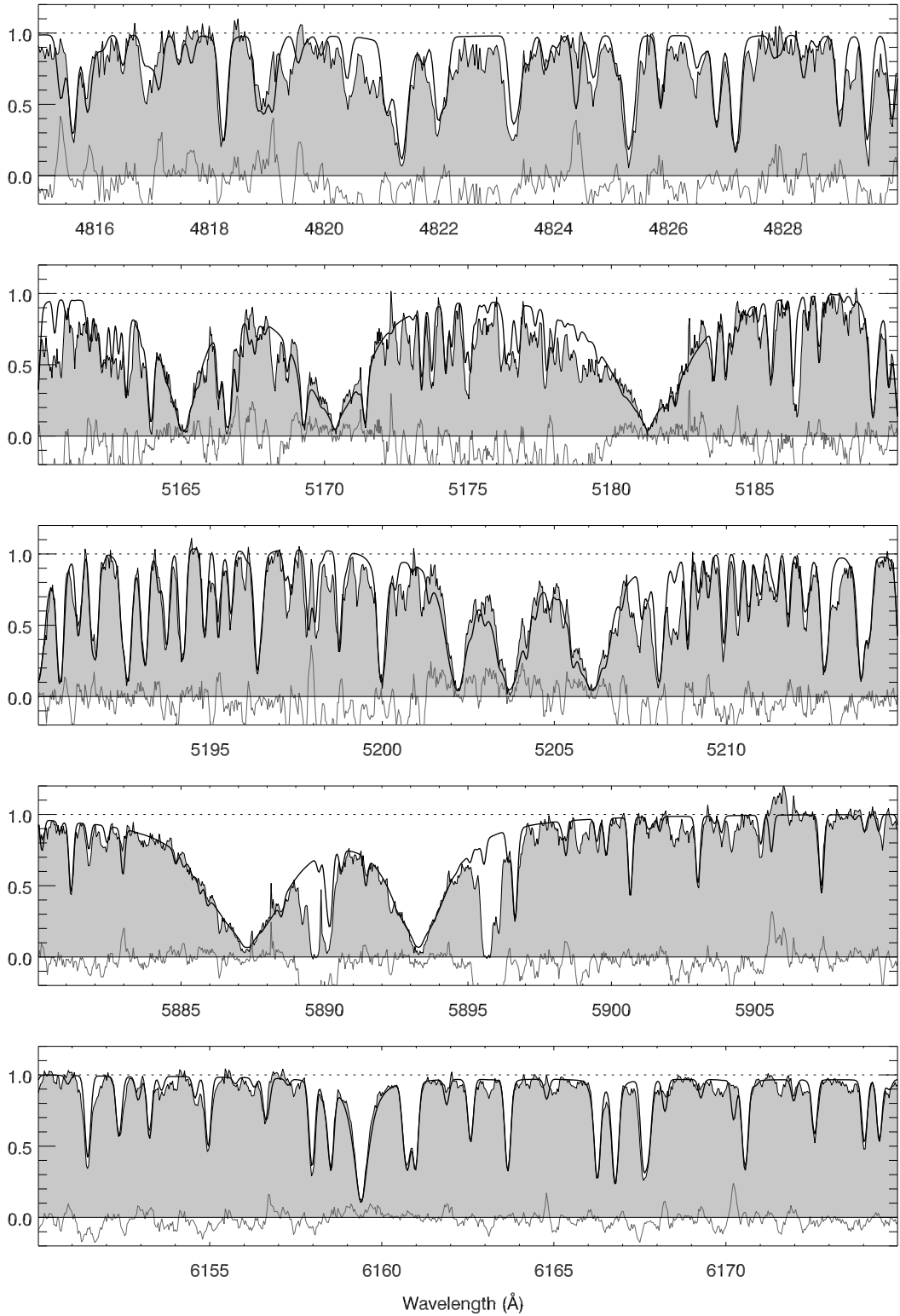}
  \caption{A comparison of the observed de-redshifted OGLE~2004--BLG--254 source star spectrum (grey)
    with a synthetic spectrum (thick black line) based on a model atmosphere
    with $\teff=4100$~K, $\logg=1.9$ and $Z=2.1\,\zsun$. The thin grey line around $y=0$ indicates the residual.
    The spectra are  normalized  to the local continuum.  From top to bottom, the panels show the
    spectral  region of: (2)~a triplet of magnesium lines; (3)~a triplet of
    chromium  lines; (4)~the NaD doublet; (1) and (5) show two other typical
    regions of  the spectrum.}
  \label{fig:spectrum}
\end{figure*}

The synthetic spectra are calculated by performing 1-dimensional radiative
transfer through standard Kurucz' model atmospheres and disk-integrating the
results from 7 $\mu$-angles. Spectral line data is taken from the {\sc VALD}
data base \citep{Kupka1999}.
The observed spectrum shows approximately $10,000$ well-defined lines, and almost
all of them are identifiable from comparison with line positions and strengths
of the transitions listed in the {\sc vald} data base.
Line profiles are computed as Voigt profiles with 
the necessary broadening parameters taken from the data base.

Among the many atomic lines, we have selected 3 particularly well
suited systems of strong \element{Mg}, \element{Cr}, and \element{Na} lines, whose intensity and line shapes are fitted to constrain the possible estimates of the 
fundamental parameters: effective temperature $\teff$, gravity $\logg$, and metallicity $Z$.
Other lines are used to control this estimate and to get a feeling for 
individual deviations from a scaled solar abundance.
\Fig{fig:spectrum} shows the observed spectrum in those regions, 
together with a synthetic spectrum discussed below.

\emph{Magnesium lines} -- The triplet of neutral magnesium lines around 
$5175$~\AA\ is well suited 
to obtain limits on the temperature and gravity.
The line system overlaps with the position of a relatively 
strong \element{MgH} band, and the ratio between \element{Mg} 
and \element{MgH} is sensitive to 
temperature as well as gravity. For strong gravity, 
the atomic magnesium triplet lines become very broad, but for
low temperatures, the balance shifts in favour 
of \element{MgH}. Therefore, the 
shape and the intensity of the atomic lines can be used together
with the ratio (or absence) of the intensity of \element{MgH} relative to 
the intensity of the atomic \element{Mg} lines to provide information on both
temperature and gravity.
The absence of \element{MgH} in our observed spectrum allows us to 
conclude that the star is not cooler than $4000$~K. The breadth 
of the atomic \element{Mg} lines allows us to confine the value of $\logg$  
between 1.0 and 2.5. 
The medium-strong neutral atomic \element{Mg} line at $5711$~\AA\ is known to respond
oppositely to the triplet lines to changes in gravity, \emph{i.e.} to become
stronger for decreasing gravity. The synthetic line is too strong for 
$\logg = 0.0$ and solar metallicity, 
while it becomes too narrow for high $\logg$. \\

\emph{Chromium lines} -- As for the \element{Mg} triplet region,
the \element{MgH} molecular system also has a relatively strong band in 
the region of a triplet of three strong chromium lines at
$5204.51$, $5206.04$ and $5208.42$ \AA, which limits 
$\teff$ to no less than $4000$~K.
Models of $\teff = 4200$~K fit the \element{Cr} lines well, while models 
of $\teff = 4400$\,K result in wings of the lines being too weak even 
for high metallicity models, while $\teff = 4000$~K would require a 
metallicity considerably below $\zsun$.
The chromium system is less sensitive to gravity than the other two
line systems, and for some $\teff$, even values as low as 
$\logg \simeq 0.0$ could be in agreement with the observed spectrum.
On the other hand the lines are sensitive to the 
chromium abundance and $Z = 0.3\ \zsun$ is too low unless $\teff$
is as low as $4000$~K. \\

\emph{Sodium lines} -- The intensity and form of the \element{NaD} 
lines around $5890 - 5896$~\AA\ 
and other neutral sodium lines are very sensitive to $\teff$ 
as well as to gravity and (sodium) abundance. Often these lines are
not useful for determination of the fundamental parameters and abundances,
because interstellar absorption saturates or changes their 
intensity. In this case, however, the main component of the 
interstellar absorption is redshifted by a velocity of $122$ km/s relative to the 
star, and the intrinsic stellar \element{NaD} lines are very strong and appear to 
be only moderately affected by interstellar absorption. The fact that the 
fundamental parameters derived from the \element{NaD} lines are in good agreement 
with the parameters derived from the other stellar lines, also 
indicates a small interstellar absorption at the radial
velocity of the star.
Model spectra from our grid with high metallicity ($Z=3\ \zsun$), high gravity ($\logg = 3$) 
or low $\teff$, all give far too
broad lines compared to the observed spectrum, and can therefore
be excluded. Models of low gravity ($\sim0$), high $\teff$ 
($\sim 4400$~K) or low $Z$ ($\sim0.3\ \zsun$) give on the other hand too 
narrow lines, and would require a
strong interstellar component at the same radial velocity as the star. \\

\begin{figure}
  \begin{center}
    \includegraphics[angle=90,width=8.5cm]{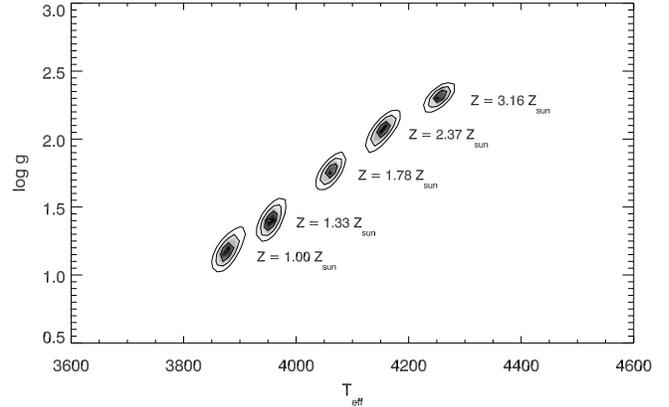}
    \caption{Using $\chi^2$ fitting of the spectrum to determine the most 
      likely values of $Z$, $\teff$ and $\logg$ reveals a strong correlation 
      between these parameters. Here we show the location of the minima in the 
      $\chi^2$ surface for five different chosen values of [\element{Fe}/\element{H}]. Around each 
      minimum we indicate the contours corresponding to confidence intervals of 
      68.3\%, 95.4\% and 99.7\% (1, 2 and 3 $\sigma$). Treating $Z$ as a free 
      parameter, we found the best $\chi^2$ solution around $\teff = 4100$~K, 
      $\logg = 1.9$ and $Z = 2.1\,\zsun$.}
    \label{fig:chi2}
  \end{center}
\end{figure}

The intensity and form of the line systems discussed above are all different
functions of $\teff$, $\logg$ and $Z$, and in principle three systems
(like the \element{Mg}, \element{Cr}, and \element{Na}
 systems) are sufficient to determine the three 
fundamental stellar parameters uniquely from the observed spectrum. 
In practice, of course, the observed spectrum is noisy, and the theoretical
spectrum suffers from inaccuracy in the model structure, incompleteness 
of the line list, etc. Therefore, rather
than a unique fit, there is a certain range of models which give 
acceptable fits to the spectrum.
If $\teff = 4300$~K is adopted, the \element{Mg} lines are well reproduced
by $\logg$ = 2 and $Z = \zsun$, while the \element{Cr} and Na lines 
and most of the other lines are slightly too weak in the wings. 
The fit to the \element{Cr} and \element{Na} lines can be improved by decreasing the 
effective temperature to 4200~K, although this slightly decreases the goodness 
of the fit to the \element{Mg} lines. A similar effect can be obtained by 
keeping the value $\teff = 4300$~K, but increasing
$\logg$ a bit, for example to $\logg = 2.25$, and increasing the metallicity.
At $\teff = 4400$~K, the fit to the \element{Mg} lines would still be correct, but 
the fit to the \element{Cr} and \element{Na} lines would be worse than for $\teff = 4300$~K,
and a compensation by increasing the gravity or the metallicity would 
require larger values than for $\teff = 4300$~K, and therefore result in 
a larger decrease in the goodness of the fit to the \element{Mg} system than for the 
$\teff = 4300$~K model. We therefore conclude that 
there is no consistent solution to the fit for $\teff = 4400$~K, and this
value is therefore too large, independently of the adopted values of 
$\logg$ and $Z$. On the other hand, changing the temperature in the other direction, for 
example to $\teff = 4000$~K, a good fit to the spectrum would require that
the value of the gravity and the metallicity be decreased.

We conclude that good fits to the \element{Mg}, \element{Cr}, and \element{Na}
lines (and other features in the spectrum)  can be obtained for models
with $\teff$ ranging from $4000$~K to $4300$~K, $\logg$ from $1.5$ to $2.5$
and $Z$ from $1.5\,\zsun$ to $2.5\,\zsun$, which classify the star as a normal K3~III red giant.
We determine the parameter values that agree best with the observed
spectrum by performing a multi-dimensional $\chi^2$-minimization using our grid
of synthetic spectra covering the range of possible valus of $\teff$, $\logg$
and $Z$. 
The optimal solution 
has $\chi^2/1504 = 2.997$,  $\teff=4100$ K, $\logg=1.9$, and $Z=2.1\,\zsun$, with a covariance of
$\teff/\logg=0.787$ and correlation coefficients of  $\teff/\logg=0.870$,
$Z/\logg=0.679$, and $Z/\teff=0.811$. Hence, we further conclude that there is a strong correlation between $\teff$,
$\logg$ and $Z$ in such a way that increasing $\teff$ must be followed by a
increased value of $\logg$ and $Z$ and vice versa.
A synthetic spectrum with the optimal parameters
is shown, together with the observed spectrum, in \Fig{fig:spectrum}. In
addition, we show as an illustration of the strong correlation between
parameters in \Fig{fig:chi2} a contour plot of the $\chi^2$-surface 
for different values of $Z$ in the plane of ($\teff$, $\logg$).

\subsection{Source radius and distance by combining spectroscopy and photometry} \label{sub:srcepos}

In general terms, photometry has a major advantage over spectroscopy in 
a higher obtainable accuracy of the broadband colours than what is obtainable 
from integrating the spectrum (whose slope is often ill defined). With good 
transformation between colours and stellar effective temperatures, an accurate 
value of $\teff$ is therefore obtainable. However, a major difficulty with 
photometry of distant stars in the Galactic plane is a large uncertainty in the
estimated reddening. We have partly got around this problem by applying the red
giant clump method, which reduces the reddening problem to a question
of the reddening of the target relative to the red giant clump, plus a
problem of defining the clump position in an observed colour-magnitude diagram 
of the field around the target.

The spectroscopic method, on the other hand, has the advantage of being
able to simultaneously estimate the stellar effective temperature and gravity 
(and metallicity). The main disadvantage, for the present purpose, is an 
ambiguity (i.e., correlation) between the obtained values of $\teff$ and 
$\logg$. We have illustrated this correlation in \Fig{fig:chi2}
for the present spectrum. By demanding that our estimated stellar
effective temperature satisfy both photometry and spectroscopy, 
we take advantage of more information than from one of the methods alone. The 
value of $\logg$ can then be restricted further than was possible from 
spectroscopy alone, by restricting the estimate of $\logg$ to the part of 
\Fig{fig:chi2} that overlaps with the photometric estimate of $\teff$. 

By combining the previous photometry and spectroscopy analyses, we get a very small 
overlap between the two ranges of effective temperature values,
namely $\teff$ in the interval $4250 - 4300$\,K, although
it is probably realistic to enlarge this estimated interval a bit.

\subsubsection{Source distance}

The combination of the fundamental source star parameters with the theoretical isochrones
of \citet{Girardi2002} and our calibrated colour-magnitude diagram can provide a measure
of the source star distance $\DS$. For a given range of $\teff$, $\logg$ and $Z$, we get from
\citet{Girardi2002} isochrones the corresponding values of the source absolute magnitude $M_I$; the 
distance then comes from $M_I=\Io-\mu$, where $\mu=5\,\log\,\DS/{\rm 10\,pc}$ is the distance modulus. 
We also require the source to be part of the Bulge, for
the probability of
observing a microlensing event for a source outside of the Bulge is very low. 

By assuming a temperature in the range $\teff = 4250 - 4300$~K, 
we find no solution, whatever the age, gravity or metallicity, as illustrated
in panel (a) of \Fig{fig:isocs1}.
This discrepancy between photometric and spectrocopic temperature cannot originate from
a higher magnification of the limb of the source, since the transit of the 
lens was already finished when the UVES spetra were obtained 
and the corresponding effect is weak (\SEC{sub:modelling}).
\citet{Fulbright05} also reported discrepancies for Bulge K-giants,
with systematic differences in photometric and spectrocopic temperatures. 

We then examine the case of higher temperatures ($\teff > 4300$~K). We first remark that
the larger the distance, the smaller the temperature has to be. the optimal solution
is then found for the maximum possible distance compatible with the source beeing a Bulge star,
which we assume to be $\DS=10.5$~kpc. Moreover, this compromise is better when the age of the source
is maximal, which means $\simeq 16$~Gyr in \citet{Girardi2002} isochrones. A possible blend of the source 
star by the lens would also imply even more discrepant results, and following \SEC{sub:modelling}, we choose to neglect it.
We study now the results involving different metallicities.
By setting for the source the red clump metallicity, $Z \simeq 0.008$ (\SEC{sec:CMDcalib}), 
we obtain a solution with $\teff \simeq 4575$~K and $\logg \simeq 2.36$, and a source
at the distance $\DS\simeq10.5$~kpc.
Although  $\teff$ is still compatible with our photometric analysis of \SEC{sec:CMDcalib},
it is above the higher bound of the allowed spectroscopic range of temperatures. The
corresponding solution is shown in panel (b) of \Fig{fig:isocs1}.
If we adopt a solar metallicity, then our requirement to have $\DS<10.5$~kpc leads
to a best solution with $\teff \simeq 4620$~K, $\logg \simeq 2.31$ with $\DS=10.5$~kpc, which 
implies similar conclusions than when assuming the red clump metallicity. This case
is shown in panel (c) of \Fig{fig:isocs1}.
We do not find any satisfying solution for higher metallicity.

In conclusion, we adopt the maximum distance compatible with the source star being
part of the Bulge, $\DS\simeq10.5$~kpc, as a best compromise between
photometric and spectroscopic analysis.

\subsubsection{Physical radius of the source} \label{sec:Sradius}

The value of the source distance $\DS$ derived above combined with the angular size of the 
source $\ThS$ is now used to determine the physical radius of the source, $\RS = \ThS\times\DS$.

The surface brightness relation from 
\citet{Kervella2004}\footnote{Although this calibration concerns
Cepheids, it has been repeatedly demonstrated that surface-brightness
relations for stable giants and Cepheids agree within $1\%$ \citep{Nordgren02}. A
possible alternative would be to make use of surface-brightness
relations directly calibrated for giant stars, but they only exist in
$(V-K)$. Such a recent calibration by \citet{Groenewegen2004} leads to
an angular radius of $4.4 \,\mu\mbox{as}$, in good agreement with our
adopted value.}:
\begin{equation} 
  \log (\ThS/\mu\mbox{as}) = 3.212 - 0.2 \,\Io + 0.421 \,\VIo
  \label{angrad}
\end{equation}
links the apparent angular radius of the source to the measured dereddened magnitude and colour 
of the source star. Adopting the values of \SEC{sec:CMDcalib}, we find an 
apparent angular radius of $\ThS = 4.5 \pm 0.7 \,\mu\mbox{as}$.

If we include a possible difference in reddening between the
source and the red giant clump, we can write the real dereddened
magnitude and colour of the source as:
\begin{eqnarray}
  \Io & = & 15.18 \pm 0.18 - \Ri \,\Delta E(V-I), \\
  \VIo & = & 1.13 \pm 0.11 - \Delta E(V-I),
\end{eqnarray}
where $\Delta$ is in the sense source minus red clump. From this, we get
an apparent angular radius:
\begin{equation}
  \log (\ThS/\mu\mbox{as}) = 0.652 - k\,\Delta E(V-I),
\end{equation}
where the coefficient of the differential extinction $k\equiv0.421-0.2\,\Ri$ 
reads $0.17 \pm 0.05$ for our adopted value of $\Ri$. 
However, a differential reddening is unlikely, since it would
imply a smaller reddening for a star behind the red clump, which is somehow unsatisfactory.

\begin{figure}
  \begin{center}
    \includegraphics[width=8.5cm]{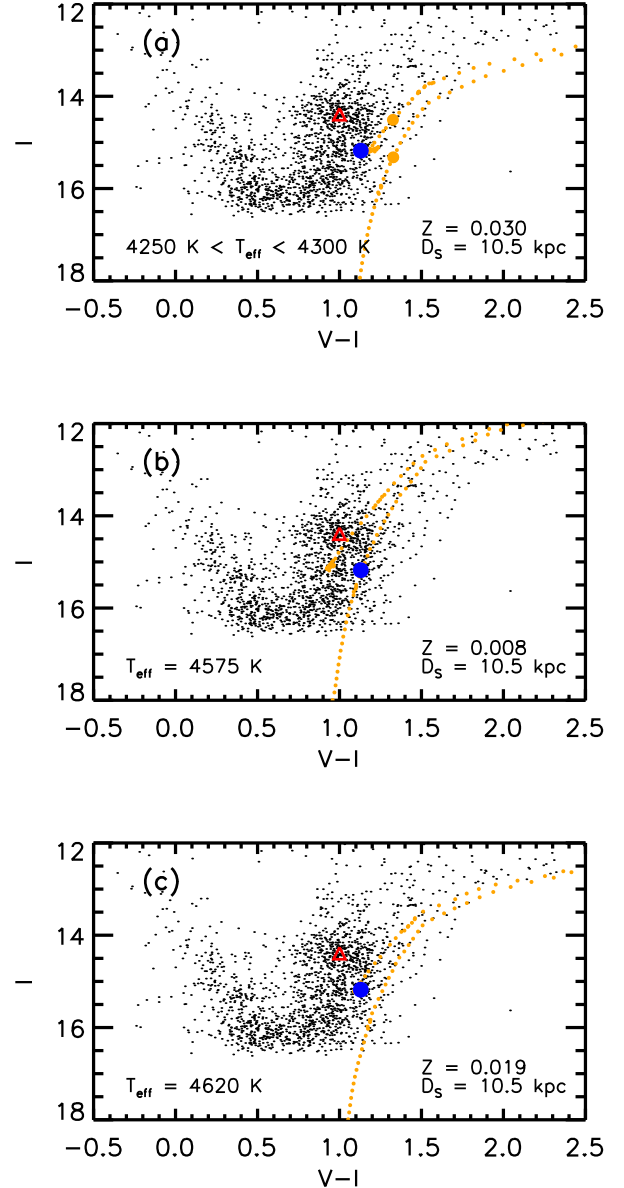}
    \caption{Same colour-magnitude diagrams as in \Fig{fig:colmag}. The de-reddened red clump is the red triangle,
      the source is the blue dot. Isochrones with metallicity $Z$ given in the right corner of each figure 
      are plotted in light orange. Photometric temperature $\teff$ and 
      source distance 
      are also indicated. An agreement between photometric and spectroscopic observations 
      is obtained when the source
      lies on the isochrone, the vertical position of which determines the source distance.
      In panel (a), it is seen that by setting the  range of temperatures
      $\teff = 4250 - 4300$~K from the spectroscopy analysis, 
      no satisfying solution can be found. In fact, the two bigger dots  on the isochrones
      matching this criterion would require a redder source and a distance which would
      put the source outside of the Bulge.
      Moreover, since the absolute magnitude of the target already agues for a source 
      on the far end of the Bulge, a differential magnification of the source relative 
      to the clump would require the isochrone to be at the left side of the blue dot,
      which would lead to even higher distances. 
      If we allow a higher temperature, solutions can be found with our requirement that $\DS<10.5$~kpc.
      Panel (b) shows the solution with the derived clump metallicity ($Z=0.008$), and 
      panel (c) with a solar metallicity. No solution is found for higher metallicities. }
    \label{fig:isocs1}
  \end{center}
\end{figure}

Thus, assuming no differential extinction, we derive a value of
$\RS \simeq 10.2\,\rsun$. We also checked that by using the simple relation $L\propto \RS^2\times T^4$,
where $T$ and $L$ are deduced from the isochrones for a source star
fulfilling the conditions shown in panel (b) or (c) of \Fig{fig:isocs1}, we find a similar 
result ($\pm2\%$) for the source radius.

\section{The lens location} \label{sec:const}

With the angular Einstein radius being related to the angular source radius $\ThS$ 
as $\ThE = \ThS/\rhoS$, we find $\ThE \simeq 112\,\mu$as. This enables us to calculate the 
the relative lens-source angular proper motion: $\mu = \ThE/\tE \simeq 3.1$~mas/yr.
The relative velocity $v$ between lens and source at the lens distance then follows from $v = \DL\,\mu$.

In principle, a measurement of the source size both in Einstein radius and physical units, as well
as the measurement of parallax parameters completely determine the lens location (given the source distnce $\DS$).

The source-size parameter $\rhoS$ (in Einstein units) is well-constrained by our 
photometric model (\SEC{sub:modelling}), and from the value of $\RS$ 
derived in \SEC{sec:Sradius}, we obtain a constraint on the lens mass $M$ as \citep{Dominik1998}:
\begin{equation}
  \frac{M}{\msun}(x) = \frac{{c^2}}{4\,G\msun}\frac{1}{\DS}\frac{\RS^2}{\rhoS^2}\frac{x}{1-x},
\label{eq:mass_x}
\end{equation}
where $x=\DL/\DS$.

\begin{figure}[!h]
 \begin{center}
  \includegraphics[width=8.5cm]{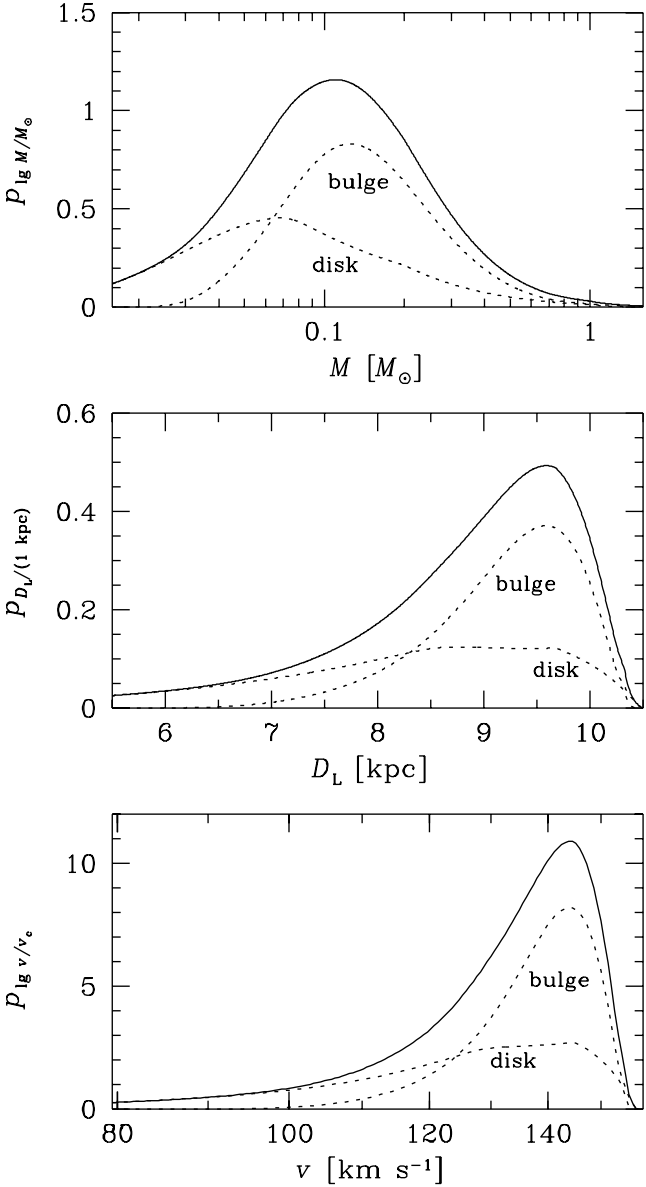}
  \caption{Probability densities for the lens mass $M$, the lens distance
$\DL$ and the relative transverse velocity $v$ at the lens
distance, assuming a source distance of $\DS = 10.5~\mbox{kpc}$ and
the adopted Galaxy model.}
  \label{fig:properties}
 \end{center}
\end{figure}

Similarly to \EQ{eq:mass_x}, a measured parallax parameter would
provide a relation between lens mass and $x$, but unfortunately the
event is too short ($\sim 13$ days $\ll$ 1 year) to provide a
measurement of the parallax, or even to give reasonable limits: values
as different as $\piE \sim 0.01$ and $10$ are hardly distinguishable
from the light curve fit. For a very similar event (duration $\tE
\simeq 13$ days, source size $\rhoS \simeq 0.06$ and very small impact
parameter $u_{\circ}$), \cite{Yoo2004} only obtained a marginal
parallax measurement too.

We then use estimates of the physical parameters, following \citet{Dominik2006} and 
assuming his adopted Galaxy model. The event time-scale $\tE = 13.23~\mbox{days}$ and the source-size parameter
$\rhoS = 0.04$ provide us with probability densities for the lens
mass $M$, the lens distance $\DL$, and the relative transverse 
velocity $v$ at the lens distance, which are shown in
\Fig{fig:properties}.

From these, we find a lens mass $M \simeq 0.11\,\msun$, a velocity $v \simeq 145~\mbox{km}\,\mbox{s}^{-1}$, and
a lens distance $\DL \simeq 9.6~\mbox{kpc}$ for an assumed source
distance of $\DS = 10.5~\mbox{kpc}$. 
The lens is preferred to reside in the Bulge, with $58\,\%$ probability, even though with a source 
on the far end of the Bulge, disk lenses play a prominent
role with $42\,\%$ probabilty for a disk lens scenario.

\begin{figure}[!h]
 \begin{center}
   \includegraphics[width=8.5cm]{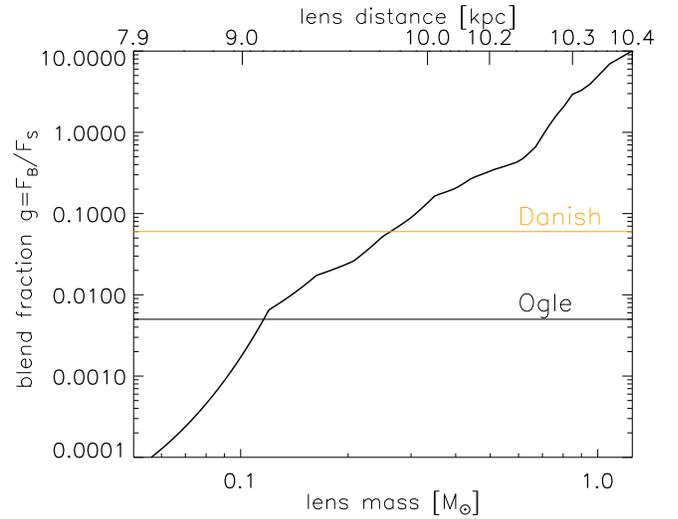}
  \caption{Assuming the fraction of blended light being solely due to the lens, upper
    limits on the lens mass can be derived from the measured blend ratio $\FB/\FS$.
    The implied blend ratio for range of main sequence lens stars of
    spectral types M0--A9 is plotted as a function of the lens distance (lower horizontal axis), 
    or equivalently as a function of the lens mass (upper horizontal axis) given \EQ{eq:mass_x}.
    The derived values are compatible with the M dwarf lens of mass 
    $\sim 0.05 - 0.3~M_\odot$.}
  \label{fig:blend}
 \end{center}
\end{figure}

Finally, similarly to the discussion of event OGLE~2002-BLG-069 \citep{Kubas2005}, 
we put upper limits on the lens mass based on the 
measured blend ratio $\FB/\FS$, assuming a luminous main-sequence lens star and a source at 
$\DS \simeq 10.5$~kpc. The latter limits, presented in \Fig{fig:blend},
are compatible with the lens being an M dwarf in the mass regime
derived in \Fig{fig:properties}.

\section{Limb-darkening measurements} \label{sub:LDmesure}

\begin{figure}[!h]
  \begin{center}
    \includegraphics[width=9cm]{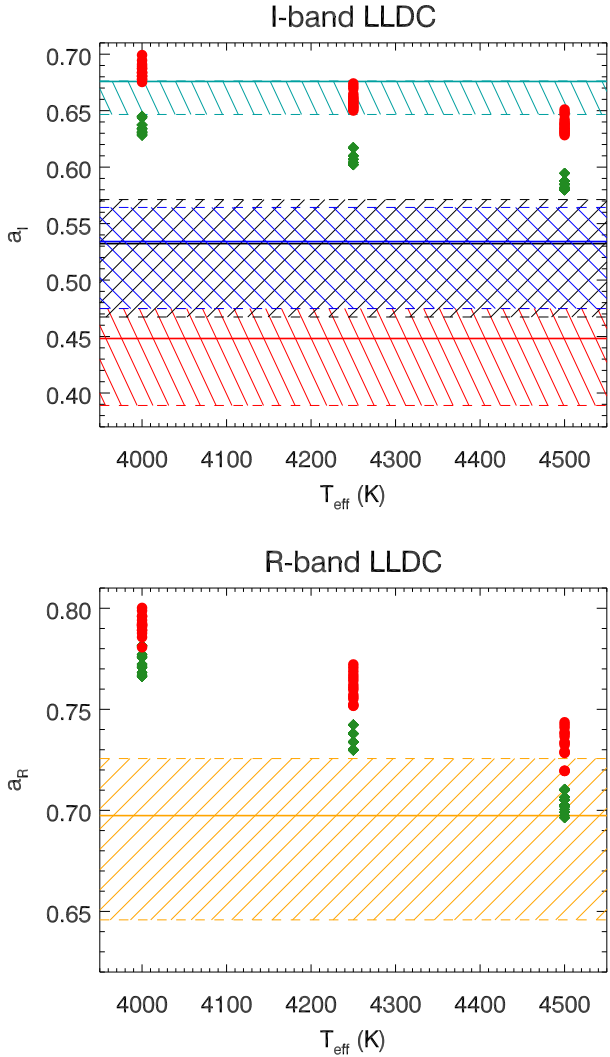}
    \caption{
      Linear limb-darkening coefficients (LLDC) for the I-band data sets (upper panel, from bottom to top: SAAO~1.0m, OGLE, UTas~1m and Boyden~1.5m) and the R-band
      data set (lower panel, Danish~1.54m). The colors follow the same convention as for \Fig{fig:lc}. 
      The best value for each $\Gamma$-coefficient is indicated by an horizontal line, where the filled area
      enclose the one-$\sigma$ error bar for each parameter.
      For the I-band, UTas~1m and OGLE give consistent values, while SAAO~1.0m and Boyden~1.5m
      are well below or above the latter ones (see text for a detailed 
      discussion). Model atmosphere LLDCs: filled red dots correspond to the Claret (2000) LLDCs while the green squares correspond
      to the  ``new fit'' LLDCs (see text), both for a range of temperature,
      $\teff=4000$, $4250$, $4500$~K (in abscissa), $logg=1.5$, $2.0$, $2.5$ and 
      $[\element{Fe}/\element{H}]=0.0$, $0.1$, $0.2$, $0.3$, $0.5$, compatible with OGLE~2004--BLG--254
      stellar parameters.}
    \label{fig:LDcompa254}
  \end{center}
\end{figure}

The photometric data of OGLE~2004--BLG--254 were dense enough to measure
limb darkening of the source. 
Because of the duration of the event, the data coverage from a single site is not sufficient for such a measurement. 
Here, we benefit from our round-the-clock follow-up that permits monitoring the event over the full course of the caustic passage. 
We find that a square root limb-darkening law 
$I(r) = \frac{1}{\pi} \left[ 1 - \Gamma \left(1-\frac{3}{2}\ \sqrt{1-r^2}\right) - \Lambda
 \left(1-\frac{5}{4}\ (1-r^2)^{1/4}\right) \right]$ does not improve the fit, and the strong 
correlation between $\Gamma$ and $\Lambda$ leads to an unsatisfactory and ambiguous result.

The limb-darkening coefficients (hereafter LLDC) derived from our best fitting model for the I- and R-band 
are presented in \Tab{tab:fitparameters} and \Fig{fig:LDcompa254}. 
For each data set, the best fitting value
is plotted as an horizontal line, within the a filled area that encompasses the allowed values 
at the one-$\sigma$ error level.
The color conventions are identical to those of \Fig{fig:lc}. We first note the very good agreement between the I-band
OGLE and UTas~1m LLDC measurements, while SAAO~1.0m and Boyden~1.5m are not consistent with the latter values.
We justify this by the fact that the peak region best constrains the LLDC, especially the shape of the top of the
light curve. As a matter of fact, UTas~1m shows a good coverage of this region, as well as OGLE and Danish~1.54m, 
as seen in \Fig{fig:lc}. However, even though SAAO~1.0m should also give good constraints, a high scatter in the peak region 
ruins the reliability of the measurement, and Boyden~1.5m exhibits large scater and has 
no baseline data points. Thus we posit
the LLDC coming from UTas~1m and OGLE (for the I-band) and Danish~1.54m (R-band) to be precise enough
to be compared with atmosphere model LLDC. As a result, the LLDC values we find for the two considered bands 
are in disagreement with the atmosphere model. The discrepancy is reduced when higher temperatures are considered, 
which tends to reinforce the conclusion from stellar parameters analysis in \SEC{sub:srceparam}.

\Fig{fig:LDcompa254} then illustrates the comparison between LLDC ATLAS atmosphere models 
predictions with OGLE~2004--BLG--254 measurements.
The filled red dots in \Fig{fig:LDcompa254} correspond to the LLDCs predicted by Claret (2000), for a range of paramater values, \emph{i.e.}
$\teff=4000$, $4250$, $4500$~K (in abscissa), $logg=1.5$, $2.0$, $2.5$ and $[\element{Fe}/\element{H}]=0.0$, $0.1$, $0.2$, $0.3$, $0.5$,   
compatible with the stellar parameters of OGLE~2004--BLG--254.
However, we point out that Claret's values are obtained by least-squares fits of intensity points more or less regularly spaced
in incidence angle $\mu=\sqrt{1-r^2}$. In terms of the radial position on the stellar disk $r$, such a fit gives very
high weight to points close to the limb and very low weight to points close to the center. In order to avoid this bias,
we interpolate the ATLAS points by cubic splines and perform the least-squares fits on the obtained 
$I(r)$ intensity profiles \citep{Heyr06}. These new LLDC values are plotted as green squares in \Fig{fig:LDcompa254}.
As shown in the last column of \Tab{tab:LDevents} for the given sources, the resulting LLDCs are systematically lower than
Claret's, in this parameter range by about $0.05$. As seen from \Fig{fig:LDcompa254}, such a difference in the LLDC may correspond 
for example to a temperature difference 
of several hundred K. 

Previous microlensing events have yielded nine limb-darkening measurements by several follow-up teams:
six for GK giants \citep{Albrow1999b,Albrow2000b,Fields2003,Cassan2004,Jiang2004,Yoo2004}, one
for a subgiant \citep{Albrow2001a}, and two for main sequence stars \citep{Afonso2000,Abe2003}.
The $I$-band GK giant results are the most relevant for comparison, and are summarized in \Tab{tab:LDevents}, 
giving the reported LLDC alongside values
derived from ATLAS models \citep{Claret2000} for comparison. 

\begin{table*}
 \begin{center}
    \begin{tabular}{l|cccc|c|c|c}
      \hline 
      \multicolumn{1}{c|}{Event} & \multicolumn{4}{c|}{Source characteristics} & Measured LLDC & ATLAS LLDC & ATLAS\\
      & Type  & $\mathrm{T_{eff}}$ & $\logg$ & [\element{Fe}/\element{H}] & $a_I$ & $a_I$ & $a_I$ (new fit) \\
      \hline\hline
      &&&&&&& \\
      MACHO~1997--BLG--28  & K2~III & 4250~K & $2.0$ & $0.0$    & $0.83\,(\pm15\,\%)$ & $0.65$ & $0.60$ \\
      MACHO~1997--BLG--41  & G5--8~III & 5000~K & $3.2$ & $-0.2$ & $0.46$ & $0.58$   & $0.53$ \\
      EROS~2000--BLG--5    & K3~III & 4200~K & $2.3$ & $+0.3$ & $0.54$ & $0.67$   & $0.62$ \\
      OGLE~2002--BLG--069  & G5~III & 5000~K & $2.5$ & $-0.3$ & $0.60$ & $0.57$   & 0.52 \\
      OGLE~2003--BLG--262  & K1--2~III & 4500~K & $2.0$ & $0.0$  &  $0.70\pm0.13$ & $0.63$ & $0.58$ \\
      OGLE~2003--BLG--238  & K2~III & 4400~K & $2.0$ & $0.0$  & $0.57\pm0.06$ & $0.64$  & $0.59$ \\
      OGLE~2004--BLG--254  & K3~III & $4000-4300$~K & $1.5$--$2.5$  & $+0.2$--$+0.4$  & $0.53~^{+0.04}_{-0.06}$ & 0.63 (closer) & 0.58 (closer) \\
      &&&&&&& \\
      \hline
    \end{tabular}
  \caption{Limb-darkening coefficients for the $I$-band of GK Bulge giants 
in OGLE~2004--BLG--254 and published events. The column ATLAS LLDC is from Claret (2000), whereas the last column is 
a new fit to ATLAS model atmosphere intensities. The best measured LLDC for OGLE~2004--BLG--254 correspond to
$\teff=4500$~K, $\logg=1.5-2.5$ and [\element{Fe}/\element{H}]$=0.-0.1$, based on the 
LLDC derived from UTas and OGLE data (see text). The ATLAS values given here
are the closest to the measured ones (\cf \Fig{fig:LDcompa254}).}
\label{tab:LDevents}
 \end{center}
\end{table*}

We exclude three cases from \Tab{tab:LDevents} for our comparison: OGLE~2002--BLG--069 because it
is of earlier spectral type; MACHO~1997--BLG--41, also of earlier type, for it suffers from 
correlations between the LLDC and other model parameters, and \object{OGLE~2002--BLG--262} which has large uncertainties due
to sparse temporal coverage. The four remaining objects are all K giant stars. The comparison between LLDC measurements
and ATLAS atmosphere models are presented in \Tab{tab:LDevents}. MACHO~1997--BLG--28, involved a cusp passage and 
seems to deviate from theory. While we do not mistrust the underlying light curve model,
the accuracy of its derived limb-darkening
coefficient is not sufficient to challenge the atmosphere modelling: an uncertainty of $15$--$20~\%$ seems
reasonable and removes part of the apparent discrepancy. 
Assuming Claret (2000) limb-darkening coefficients for the stars with their known temperature, 
the two remaining literature events, EROS~2000--BLG--5 and \object{OGLE~2003--BLG--238} -- together with OGLE~2004--BLG--254 --,
appear to disagree with theory predictions, 
considering the high quality and suitability of the data for a direct surface-brightness profile constraint.
However, assuming our ``new fit'' limb-darkening coefficients, OGLE~2003--BLG--238 now agrees with the prediction, 
EROS~2000--BLG--5 is much closer to the predicted model, and the discrepancy for OGLE~2004--BLG--254 is reduced.

There is still an incompatibility between
the measured and theoretical limb darkening of Bulge K-giant stars, as previously for EROS~2000--BLG--5 \citep{Fields2003}. 
However, if a limb-darkening coefficient following our fitting prescription is used, the discrepancy 
is significantly smaller than with the coefficients published by Claret (2000).

Finally, a comparison between the shape of synthetic atmosphere models and linear and square-root limb-darkening
law curves suggests that the classical laws are too restrictive to fit well the microlensing observations, 
as already suggested by \citet{Heyrovsky2003}.
For example, all the normalized curves derived from the classical laws are constrained to pass through the point at
$r=\sqrt{5}/3$, whereas curves derived from synthetic spectra of giant stars tend to intersect $\sim5$~\%
closer to the center. If the latter is closer to reality, the LDC inferred from a given microlensing event might be biased by
the attempt to compensate for the too steep outer behaviour of the linear limb-darkening law,
and this would only be exacerbated by any sparsity in the photometric coverage
at that phase. 
We note here that \citet{Heyrovsky2003} using simulated single-lens microlensing
light curves showed that the accuracy of recovering linear limb-darkening coefficients is limited
by the inadequacy of the linear limb-darkening law. \citet{Heyrovsky2003} suggests addressing this problem using a 
principal component analysis (PCA) approach,
where orthogonal basis functions extracted from a grid of atmosphere models 
are used to describe the broad-band limb darkening of stars. This will be investigated in a forthcoming paper
together with a set of similar single lens events with finite source effects.

\section{Summary and conclusion} \label{sec:discussion}

We have performed dense photometric monitoring of the microlensing
event OGLE~2004--BLG--254, a relatively short duration and small
impact parameter microlensing event generated by a point-like lens
transiting a giant star. The peak magnification was about $A_{\circ}\sim 60$,
effectively multiplying the diameters of our network
telescopes by a factor $\sim 8$.  High-resolution spectra were taken
using the UVES spectrograph at La Silla while the source was magnified
by a factor $A \sim 20$, just after the end of the transit of the
source over the caustic. This yielded a precise measurement of the
characteristics of the star, a K3~III giant, although a discrepancy
is found between photometric and spectroscopic temperature.
Using a calibrated colour-magnitude diagram analysis and isochrones, 
we find a source angular radius $\ThS \simeq 4.5 \pm 0.7 \,\mu\mbox{as}$, 
and a physical
radius $\RS \simeq 10.2 \rsun$ with the source distance $\DS = 10.5$~kpc.
A Galaxy model
together with the event time-scale $\tE = 13.23~\mbox{days}$ and the
source-size parameter $\rhoS = 0.04$ yielded a lens mass $M\simeq 0.11\,\msun$, 
a lens distance $\DL = 9.6~\mbox{kpc}$, and a
velocity $v \simeq 145~\mbox{km}\,\mbox{s}^{-1}$ at the lens distance.
From our photometric data, we have derived
measurements of the source limb-darkening coefficients for the $I$ and
$R$ broadband filters, and provided arguments for a discussion about a
lack of relevant physics in K-giants atmosphere models, 
or an inadequacy of the linear limb-darkening law to model. 

\begin{acknowledgements}

We express our gratitude to the ESO staff at Paranal
for reacting at short notice to our ToO request.
We are very grateful  to the observatories that support our
science (European Southern Observatory, Canopus, CTIO, Perth, SAAO)
via the generous allocations of time that make this work possible. 
The operation of Canopus Observatory is in part supported by a financial
contribution from David Warren, and the Danish telescope at La Silla
is operated by IDA financed by SNF. 
Partial support to the OGLE project was provided by the following
grants: Polish MNSW BW grant for Warsaw University Observatory, 
NSF grant AST-0204908 and NASA grant NAG5-12212.
This publication makes use of data products from the 2MASS and DENIS projects, 
as well as the SIMBAD database, Aladin and Vizier catalogue operation tools
(CDS Strasbourg, France).
AC and DK acknowledge the ``EGIDE'' grant for 
Paris-Berlin travel support. MD acknowledges postdoctoral support on the PPARC rolling
grant PPA/G/O/2001/00475. DH acknowledges support by the Czech Science Foundation grant
GACR 205/04/P256. The authors are thankful for the helpful comments and suggestions
of the anonymous referee.
\end{acknowledgements}

\bibliographystyle{aa}
\bibliography{4414bibl}
\end{document}